\RequirePackage{lineno}
\documentclass[aps,prl,twocolumn,groupedaddress,amsmath,amssymb]{revtex4-1}

\usepackage{amssymb}
\usepackage{graphicx}
\usepackage{subfigure}
\usepackage{dcolumn}
\usepackage{bm}
\usepackage{epsfig}
\usepackage{xcolor}
\usepackage{enumerate}
\usepackage{booktabs}
\usepackage{hyperref}
\hypersetup{hidelinks,
	colorlinks=true,
	allcolors=blue,
	pdfstartview=Fit,
	breaklinks=true}
\usepackage{float}

\begin{document}

\newcommand{\jpsi}{J/\psi}
\newcommand{\pip}{\pi^+}
\newcommand{\pin}{\pi^-}
\newcommand{\pio}{\pi^0}
\newcommand{\g}{\gamma}
\newcommand{\gev}{GeV/c$^2$}
\newcommand{\mev}{MeV/c$^2$}
\newcommand{\ar}{\rightarrow}
\newcommand{\ks}{K_S^{0}}
\newcommand{\etap}{\eta^\prime}

\title{\Large \boldmath \bf Observation of the Anomalous Shape of $X(1840)$ in $\jpsi\rightarrow \gamma 3(\pip\pin)$ Indicating a Second Resonance Near $p\bar{p}$ Threshold}

\author{
\begin{small}
\begin{center}
M.~Ablikim$^{1}$, M.~N.~Achasov$^{5,b}$, P.~Adlarson$^{75}$, X.~C.~Ai$^{81}$, R.~Aliberti$^{36}$, A.~Amoroso$^{74A,74C}$, M.~R.~An$^{40}$, Q.~An$^{71,58}$, Y.~Bai$^{57}$, O.~Bakina$^{37}$
% ...
\\
\vspace{0.2cm}
(BESIII Collaboration)\\
\vspace{0.2cm} {\it
$^{1}$ Institute of High Energy Physics, Beijing 100049, People's Republic of China\\
% ...
$^{5}$ Budker Institute of Nuclear Physics SB RAS (BINP), Novosibirsk 630090, Russia\\
% ...
$^{36}$ Johannes Gutenberg University of Mainz, Johann-Joachim-Becher-Weg 45, D-55099 Mainz, Germany\\
$^{37}$ Joint Institute for Nuclear Research, 141980 Dubna, Moscow region, Russia\\
% ...
$^{40}$ Liaoning Normal University, Dalian 116029, People's Republic of China\\
% ...
$^{57}$ Southeast University, Nanjing 211100, People's Republic of China\\
$^{58}$ State Key Laboratory of Particle Detection and Electronics, Beijing 100049, Hefei 230026, People's Republic of China\\
% ...
$^{71}$ University of Science and Technology of China, Hefei 230026, People's Republic of China\\
% ...
$^{74}$ University of Turin and INFN, (A)University of Turin, I-10125, Turin, Italy; (B)University of Eastern Piedmont, I-15121, Alessandria, Italy; (C)INFN, I-10125, Turin, Italy\\
$^{75}$ Uppsala University, Box 516, SE-75120 Uppsala, Sweden\\
% ...
$^{81}$ Zhengzhou University, Zhengzhou 450001, People's Republic of China\\

\vspace{0.2cm}
% ...
$^{b}$ Also at the Novosibirsk State University, Novosibirsk, 630090, Russia\\
% ...

}

\end{center}
\vspace{0.4cm}
\end{small}
}


\begin{abstract}

Using a sample of $(10087\pm44)\times 10^6$ $J/\psi$ events, which is about 45 times larger than that was previously analyzed, a further investigation on the $\jpsi\rightarrow \gamma 3(\pip\pin)$ decay is performed. 
A significant distortion at 1.84 GeV/$c^2$ in the line-shape of the $3(\pip\pin)$ invariant mass spectrum is observed for the first time, which could be resolved by two overlapping resonant structures, $X(1840)$ and $X(1880)$. The new state $X(1880)$ is observed with a statistical significance larger than $10\sigma$. The mass and width of $X(1880)$ are determined to be $1882.1\pm1.7\pm0.7$ MeV/$c^2$ and  $30.7\pm5.5 \pm2.4$ MeV, respectively, which indicates the existence of a $p\bar{p}$ bound state.
\end{abstract}

\maketitle

A distinct resonance, $X(1835)$~\cite{X1835_bes_2005}, in the $\pi^+\pi^-\eta^\prime$ invariant mass spectrum and a dramatic $p\bar{p}$ mass threshold enhancement~\cite{Xppbar_bes_2003} in $J/\psi\rightarrow\gamma p\bar{p}$ were first observed by BESII, which stimulated both theoretical and experimental interests in their nature.
Some theoretical models are proposed to interpret their internal structures, e.g. a $p\bar{p}$ bound state~\cite{X1835_JP_2009,X1835_xiaohai_2009,X1835_guijun_2006,X1835_zhigang_2007}, a pseudoscalar glueball ~\cite{X1835_bingan_2006,X1835_nikolai_2006,X1835_gang_2006}, or a radial excitation of the $\etap$ meson~\cite{X1835_tao_2006}. 
Subsequently these resonances were confirmed by BESIII~\cite{Xpp_bes3_2010, X1835_bes3_2011} and CLEO~\cite{Xpp_cleo_2010} experiments and found to have the same $J^{PC}$ of $0^{-+}$~\cite{Xpp_bes3_2011, X1835_bes3_2015}.
Meanwhile, a prominent structure, $X(1840)$, was observed in the 3$(\pi^+\pi^-)$ invariant mass ($M(6\pi)$) spectrum in $J/\psi\rightarrow \gamma 3(\pi^+\pi^-)$ with a mass of $1842.2\pm4.2_{-2.6}^{+7.1}$ MeV/$c^2$ and a width of $83\pm14\pm11$ MeV~\cite{X1840_bes3_2013}. It was interpreted as a new decay mode of $X(1835)$, although its width is substantially narrower than that of $X(1835)$~\cite{X1835_bes3_2011}. 
Of interest is that an updated analysis of $J/\psi\rightarrow\gamma\pi^+\pi^-\eta^\prime$ observed a significant abrupt change in slope of the $X(1835)\rightarrow \pi^+\pi^-\eta^\prime$ line-shape at the $p\bar{p}$ mass threshold, which could be originated from the opening of an additional $p\bar{p}$ decay channel (threshold effect) or the interference between two different resonance contributions~\cite{X1835_fit}.
To understand whether a similar phenomenon to that of $J/\psi\rightarrow\gamma\pi^+\pi^-\eta^\prime$ exists around the $p\bar{p}$ mass threshold in the $M(6\pi)$ spectrum, it is worth a more detailed investigation on the $X(1840)$ line-shape in $J/\psi\rightarrow\gamma 3(\pi^+\pi^-)$ with higher precision.
In this letter we report an anomalous line-shape of $X(1840)$ in the $M(6\pi)$ spectrum in $\jpsi\ar \gamma 3(\pip\pin)$ with a sample of $(10087\pm44) \times 10^6$ $J/\psi$ events~\cite{num_of_jpsi} collected with the BESIII detector. The size of the sample is about 45 times greater than that used in Ref.~\cite{X1840_bes3_2013}.

The BESIII detector records symmetric $e^+e^-$ collisions provided by the BEPCII storage ring~\cite{bepc2} in the center-of-mass energy range from 2.0 to 4.95 GeV, which is described in detail in~\cite{BESIIIdetector, mrpc1, mrpc2, mrpc3}. Simulated data samples produced with a {\sc geant4}-based~\cite{geant4} Monte Carlo (MC) package, which includes the geometric description of the BESIII detector~\cite{Huang:2022wuo} and the detector response, are used to determine detection efficiencies and to estimate backgrounds. The simulation models the beam energy spread and initial state radiation (ISR) in the $e^+e^-$ annihilation with the generator {\sc kkmc}~\cite{kkmc}. All particle decays are modelled with {\sc evtgen}~\cite{evtgen1, evtgen2} using branching fractions either taken from the Particle Data Group~\cite{PDG}, when available, or otherwise estimated with {\sc lundcharm}~\cite{lund1, lund2}. Final state radiation~(FSR) from charged final state particles is incorporated using the {\sc photos} package~\cite{photos}.

Charged tracks detected in the main drift chamber (MDC) are required to be within a polar angle ($\theta$) range of $|\rm{cos\theta}|<0.93$, where $\theta$ is defined with respect to the $z$-axis, the symmetry axis of the MDC. The distance of closest approach to the interaction point must be less than 10\,cm along the $z$-axis, and less than 1\,cm in the transverse plane. Photon candidates are reconstructed using clusters of energy deposited in the 
electromagnetic calorimeter (EMC), where a minimum energy of 25 MeV for the barrel region ($\lvert\cos\theta\rvert<0.8$) and 50 MeV for the endcap region ($0.86<\lvert\cos\theta\rvert<0.92$) is required. To suppress electronic noise and showers unrelated to the event, the difference between the EMC time and the event start time is required to be within 
[0, 700]\,ns.

Candidates for the signal are required to have six charged tracks with zero net charge and at least one photon.
All the charged tracks are assumed to be pions.
A four-momentum-constraint (4C) kinematic fit is performed under the hypothesis of $\jpsi\ar\gamma3(\pip\pin)$, and the $\chi^2_{\rm{4C}}$ of this kinematic fit is required to be less than 30. 
For events with more than one photon candidate, the $\gamma3(\pip\pin)$ combination with the minimum $\chi^2_{\rm{4C}}$ is retained. 
To suppress the backgrounds with a final state of $\gamma\gamma3(\pip\pin)$, the $\chi^2_{\rm{4C}}$ is required to be less than that for the kinematically similar $\gamma\gamma3(\pip\pin)$ hypothesis.
Furthermore, for the events containing at least two photons, the $\gamma\gamma$ invariant mass is required to be outside the $\pi^0$ mass window of $\lvert M_{\gamma\gamma}-m_{\pi^0}\rvert<0.01$ $\mathrm{GeV}/c^2$ to veto the backgrounds with $\pi^0$ in their final states.
 
The $\jpsi\ar\g\ks\ks\pip\pin$ process with a subsequent decay of $\ks$ to $\pip\pin$ has the same final state as the signal decay. To suppress this background, the $K_S^0$ candidates are reconstructed from secondary vertex fit (SVF) to all $\pip\pin$ pairs. The $K_S^0$ candidates are tagged by passing the SVF successfully and requiring the $\pip\pin$ invariant mass in a range of $\lvert M_{\pip\pin}-m_{K_S^0}\rvert<0.005$ $\mathrm{GeV}/c^2$, where $m_{K_S^0}$ is the $K_S^0$ known mass. Given the existence of mis-reconstructed $K_S^0$ for the signal, events with the number of $K_S^0$ candidates less than 2 are retained for further analysis.

\begin{figure}
    \centering
    \includegraphics[height=5cm,width=7.cm]{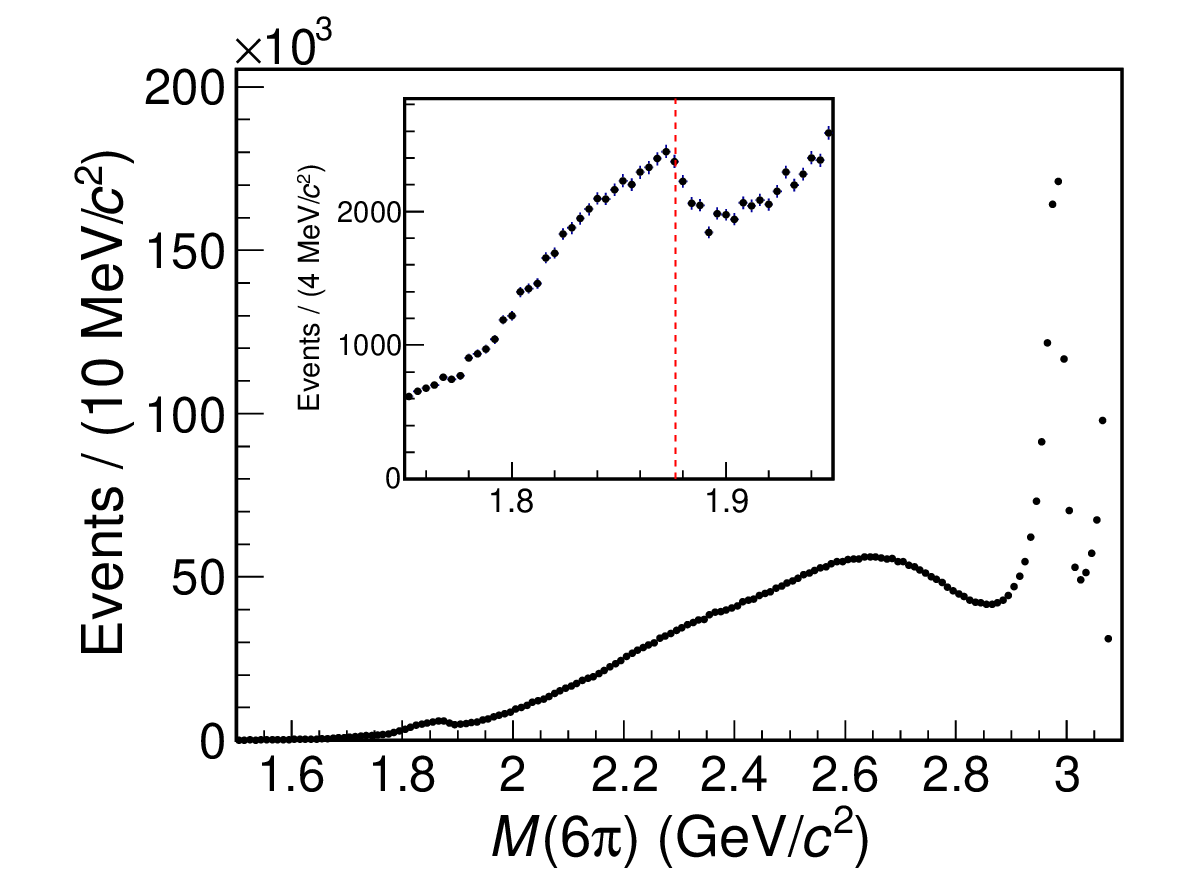}
    \caption{$M(6\pi)$ distribution from $\jpsi\ar\gamma3(\pip\pin)$ events. The dots with error bars are data. The inset shows the data between 1.75 and 1.95 GeV/$c^2$ and the vertical dotted line represents the $p\bar{p}$ mass  threshold.\label{m6pi}}
\end{figure}

After applying the above requirements, the $M(6\pi)$ spectrum is shown in Figure~\ref{m6pi}, where, in addition to the well established $\eta_c$ peak and the peak around 3.07 GeV/$c^2$ from $J/\psi \rightarrow 3(\pip\pin)$ background channel, a distinct structure around 1.84 GeV/$c^2$ is apparent, and an anomalous line-shape near the $p\bar{p}$ mass threshold is clearly observed, as shown in the inset plot. 

With exactly the same processes of simulated inclusive $J/\psi$ events as for the data, no peaking background contribution around 1.84 GeV/$c^2$ is found. 
The remaining background is mainly from $\jpsi\ar\pi^03(\pip\pin)$, for which, we use a one-dimensional data-driven method to determine its contribution. We select the $\jpsi\ar\pi^03(\pip\pin)$ events from data firstly and then implement the signal selection criteria on these events. The $M(6\pi)$ spectrum extracted based on these surviving events is further reweighted by the ratio of MC-determined efficiencies for $\jpsi\ar\g3(\pip\pin)$ to $\jpsi\ar\pi^03(\pip\pin)$ events.
To ensure that the anomalous line-shape in data is not caused by the distortion of the detection efficiency due to event selection bias, we studied the phase space MC events of $\jpsi\ar\g3(\pip\pin)$. As a result, 
neither the 1.84 GeV/$c^2$ peaking structure nor the abrupt change in the line-shape near the $p\bar{p}$ mass threshold is caused by the background processes or the distortion of the the event selection efficiency.

\begin{figure}
\centering
\includegraphics[width=3in]{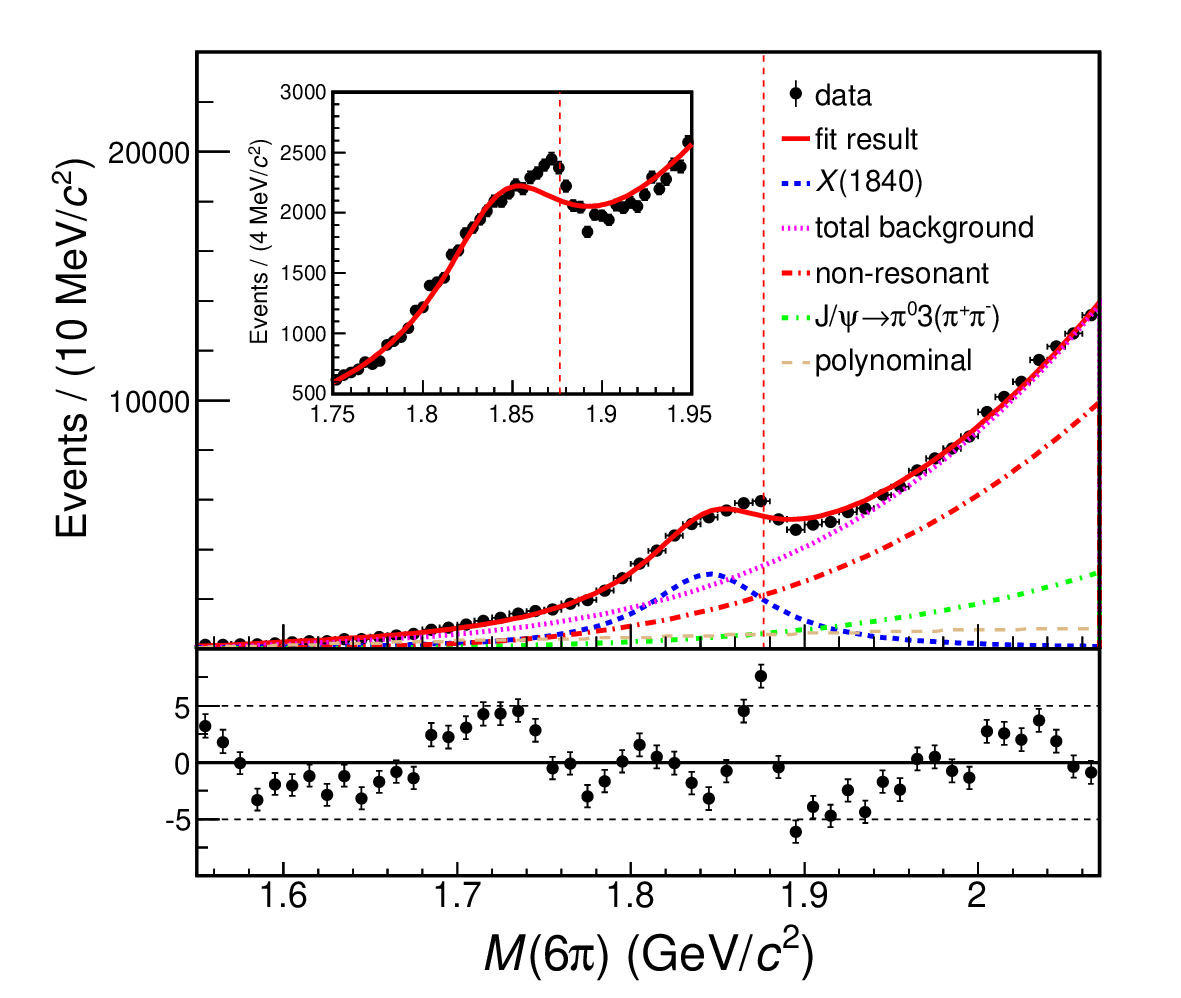} 
\caption{\label{fig_bw}Fit result of the $M(6\pi)$ distribution with a simple BW function. The dots with error bars are data, the solid curve in red is the total fit result, the dashed line in blue is the $X(1840)$ signal, the dash-dotted line in green is the background events from $\jpsi\ar\pi^0 3(\pip\pin)$, and the dotted line in magenta is the sum of background.}
\end{figure}

We perform an unbinned maximum likelihood fit to the $M(6\pi)$ spectrum between 1.55 and 2.07 GeV/$c^2$ with the $X(1840)$ peak represented by the efficiency corrected Breit-Wigner (BW) function convolved with a Gaussian function to account for the mass resolution, which is determined to be 4 MeV/$c^2$ from the MC simulation. 
The dominant background to the $X(1840)$ peak is from the non-resonant contribution of $J/\psi \rightarrow \gamma3(\pi^+\pi^-)$, whose shape is obtained through MC simulation and the fraction is free in the fit. The $J/\psi\rightarrow 3(\pi^+\pi^-)\pi^0$ background contributions are estimated with the data-driven approach as described above. The remaining background is described by a free second-order polynomial function. Without explicit mention, all components are treated as incoherent contributions. The fit quality is significantly poor, as shown in Fig.~\ref{fig_bw}. The goodness of fit is studied using a $\chi^2 $ test and the $\chi^2$ value per number of degrees of freedom ($ndof$) is found to be $\chi^2/ndof=399.0/45$. This implies that a simple resonant structure fails to describe the $M(6\pi)$ spectrum.

\begin{figure}
\centering
\subfigure{\label{fig_flatte}
\includegraphics[width=3in]{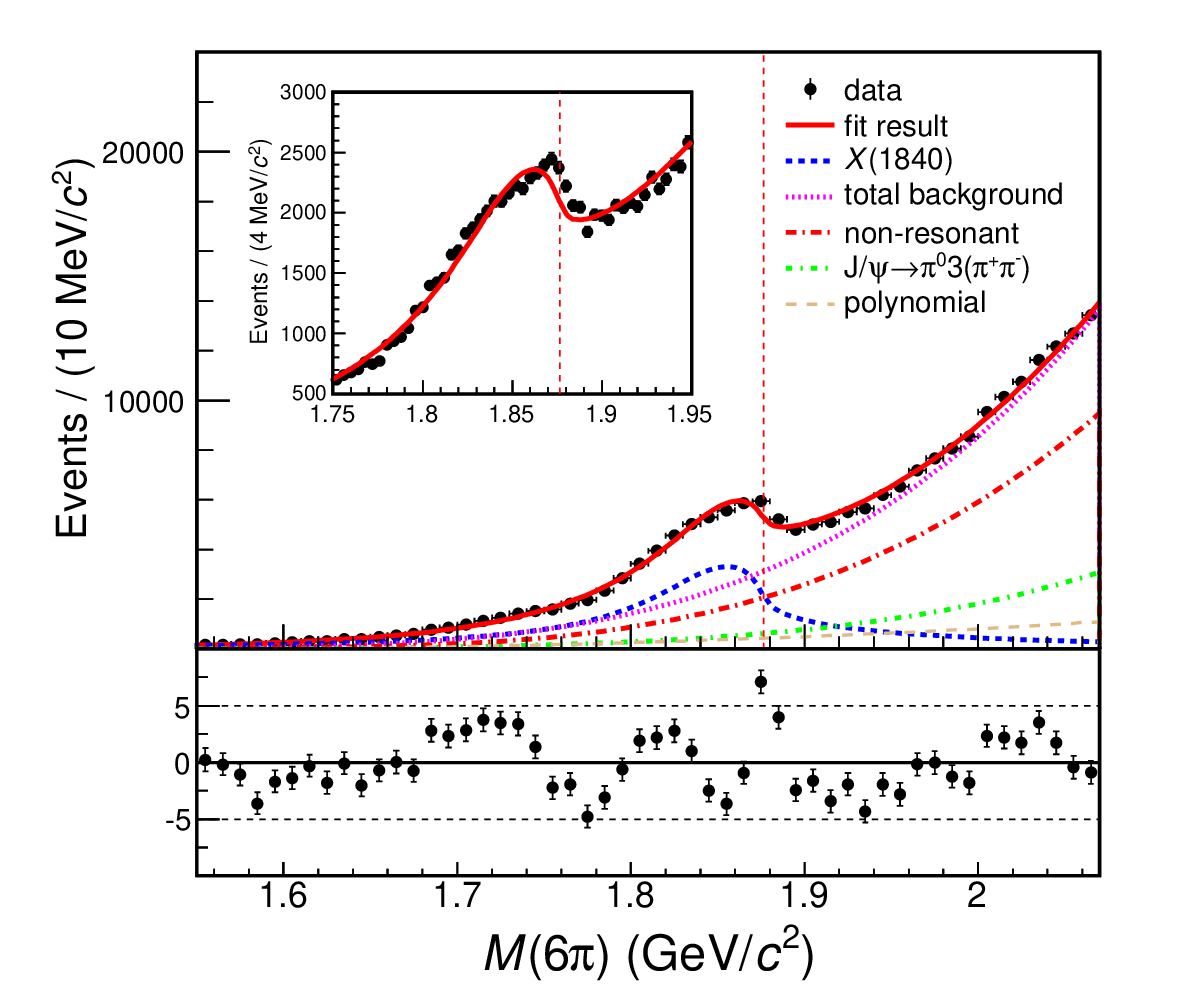} 
\put(-40,165){\bf \large~(a)}
}
\subfigure{\label{fig_bwbw}
\includegraphics[width=3in]{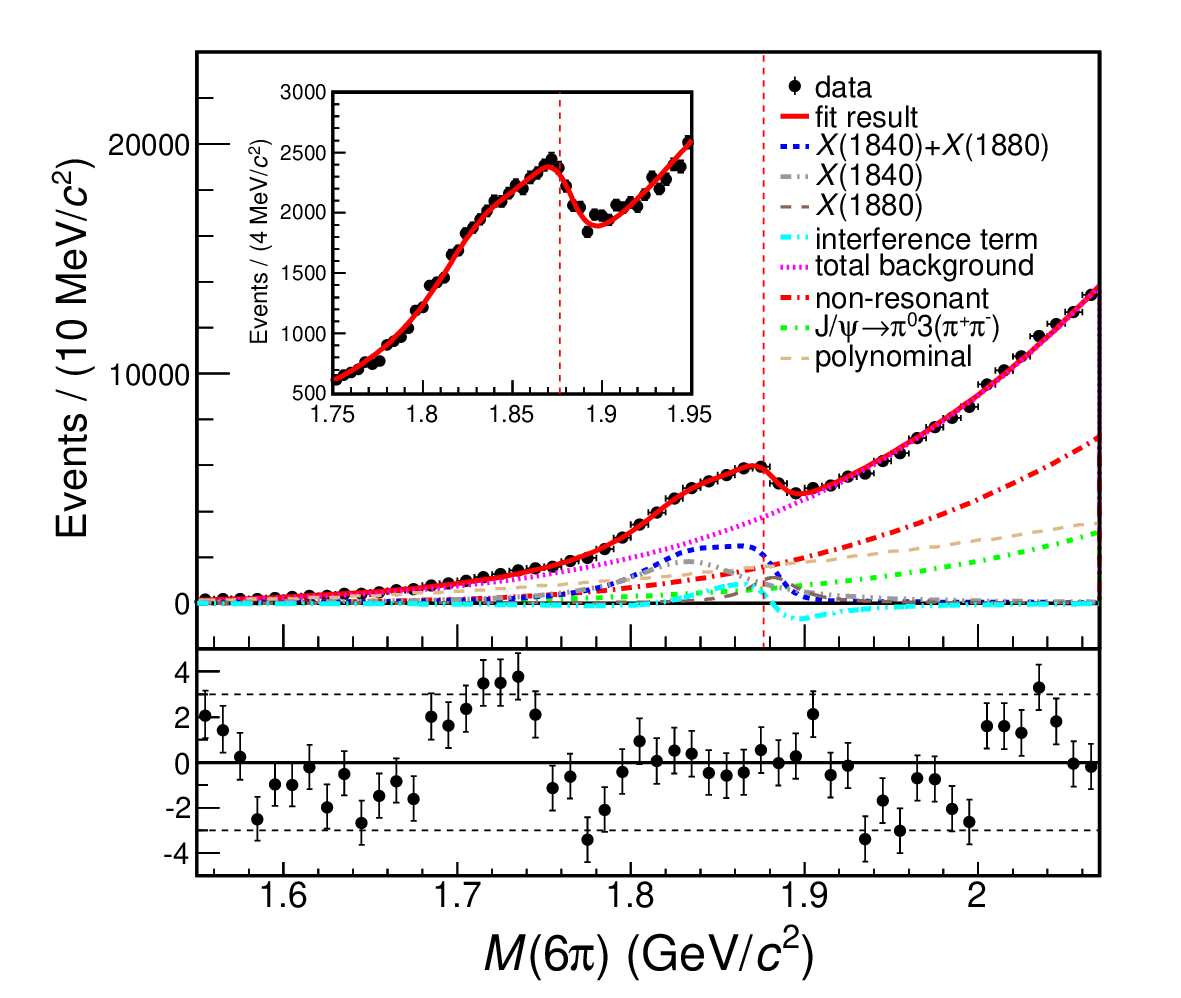}
\put(-40,165){\bf \large~(b)}
}
\caption{Fit result of the $M(6\pi)$ distribution with model {\bf I} (a), and solution I for the model {\bf II} (b). The dashed line in blue is the $X(1840)$ signal for (a), and the sum of $X(1840)$ and $X(1880)$ for (b).}
\end{figure}

To resolve the discrepancy from data, two different models for the line shape of the structure around 1.84
GeV/$c^2$ are applied to investigate the resonances in the $M(6\pi)$ spectrum.
With an assumption of  the line-shape of $3(\pip\pin)$ above the $p\bar{p}$ mass threshold affected by the opening of the $X(1840)\ar p \bar{p}$ decay (model {\bf I}),  we try to describe the anomalous shape with a Flatt$\rm{\acute{e}}$ formula~\cite{flatte},
\begin{linenomath*}
\begin{equation}
        A={\lvert\frac{1}{M^2-s-i\sum_j g_{j}^2\rho_j}\rvert}^2 \label{flatte},\nonumber
\end{equation}
\end{linenomath*}
\noindent where $M$ is a parameter with the dimension of mass, $s$ is the mass square of the $3(\pip\pin)$ combination, $\rho_j$ is the phase space for the decay mode $j$, and $g^2_j$ is the corresponding coupling strength. The $\sum_j g_{j}^2\rho_j$ term describes how the decay width varies with $s$. Approximately,
\begin{linenomath*}
\begin{equation} \label{grho}
        \sum_j g_{j}^2\rho_j\approx g^2_0(\rho_0+\frac{g^2_{p\bar{p}}}{g^2_0}\rho_{p\bar{p}}),
\end{equation}
\end{linenomath*}
\noindent where $g^2_0$ is the sum of $g^2$ of all decay modes other than $X(1840)\ar p\bar{p}$, $\rho_0$ is the maximum two-body decay phase space volume~\cite{PDG} and $g^2_{p\bar{p}}/g^2_0$ is the ratio between the coupling strength to the $p\bar{p}$ channel and the sum of all other channels.
This fit, as illustrated in Fig.~\ref{fig_flatte}, yields $M = 1.818 \pm 0.009$ GeV/$c^2$, $g^2_0 = 18.0 \pm 2.8$ ${\rm{GeV}}^2/c^4$, and $g^2_{p\bar{p}} = 51.4 \pm 14.8$ ${\rm{GeV}}^2/c^4$. This model fit has a $\text{log}~\mathcal{L}$ that is improved over the simple Breit-Wigner one by 42.8. The significance of $g^2_{p\bar{p}}/g^2_0$ being non-zero is $9.2\sigma$. The goodness of the fit $\chi^2/ndof=317.9/44$, yet not enough to be acceptable for a good description of data.

A comparison between the fit result of model {\bf I} and the data
reveals a tension around  the $p\bar{p}$ mass threshold. To obtain a better description on data, another model allows for interference between two resonant components (model {\bf II}) and the coherent sum of them is defined as
\begin{linenomath*}
\begin{equation} \label{eq_bwbw}
        A={\lvert\frac{1}{M^2_1-s-iM_1\Gamma_1} + \beta\frac{1}{M^2_2-s-iM_2\Gamma_2}\rvert}^2,
\end{equation}
\end{linenomath*}
\noindent where $M_1$, $\Gamma_1$, $M_2$ and $\Gamma_2$ represent the masses and widths of the two resonant structures, denoted as $X(1840)$ and $X(1880)$, respectively. $\beta$ is a complex parameter accounting for the contribution of $X(1880)$ relative to the $X(1840)$ as well as the phase between them.

The fit with model {\bf II} improves the fit quality significantly ($\chi^2/ndof=155.6/41$), in particular for the region around the $p\bar{p}$ mass threshold, which is illustrated in Fig.~\ref{fig_bwbw}. The $\chi^2$ value in the region of 1.78 to 1.92 GeV/$c^2$ changes from 135.6 for model {\bf I} to 15.1 for model {\bf II}, which shows the latter model provides a significant improvement in describing the anomalous line-shape. The masses, widths and signal yields of these two resonant components, as summarized in Table~\ref{table_finalresult}, are determined to be $M_{X(1840)}=1832.5\pm3.1$ $\mathrm{MeV}/c^2$, $\Gamma_{X(1840)}=80.7\pm5.2$ $\rm{MeV}$,  $N_{X(1840)}=20980\pm5341$, $M_{X(1880)}=1882.1\pm1.7$ $\mathrm{MeV}/c^2$, $\Gamma_{X(1880)}=30.7\pm5.5$ $\rm{MeV}$, $N_{X(1880)}=5460\pm3757$, where the uncertainties are statistical only. The $\text{log}~\mathcal{L}$ of this fit is improved by 73.6 over that of the fit with model {\bf I}. The statistical significance of $X(1880)$ is found to be larger than 10$\sigma$, which is determined by comparing the log-likelihood value and the number of degrees of freedom between model {\bf II} and model {\bf I} using Wilks' theorem~\cite{Wilks:1938dza}.
As discussed in Ref.~\cite{multisolutions}, a fit using a coherent sum of two BW functions may result in two nontrivial solutions with the same resonant parameters. We make an extensive investigation on the fit and find the second solution with the same fit quality, which  yields  $N_{X(1840)}=36506\pm8740$ and $N_{ X(1880)}=22097\pm5794$,  corresponding to the destructive interference as described in Ref.~\cite{multisolutions}. The figure of second solution is given in the Supplemental Material~\cite{sm}.

\begin{figure}
    \includegraphics[height=5cm,width=7.cm]{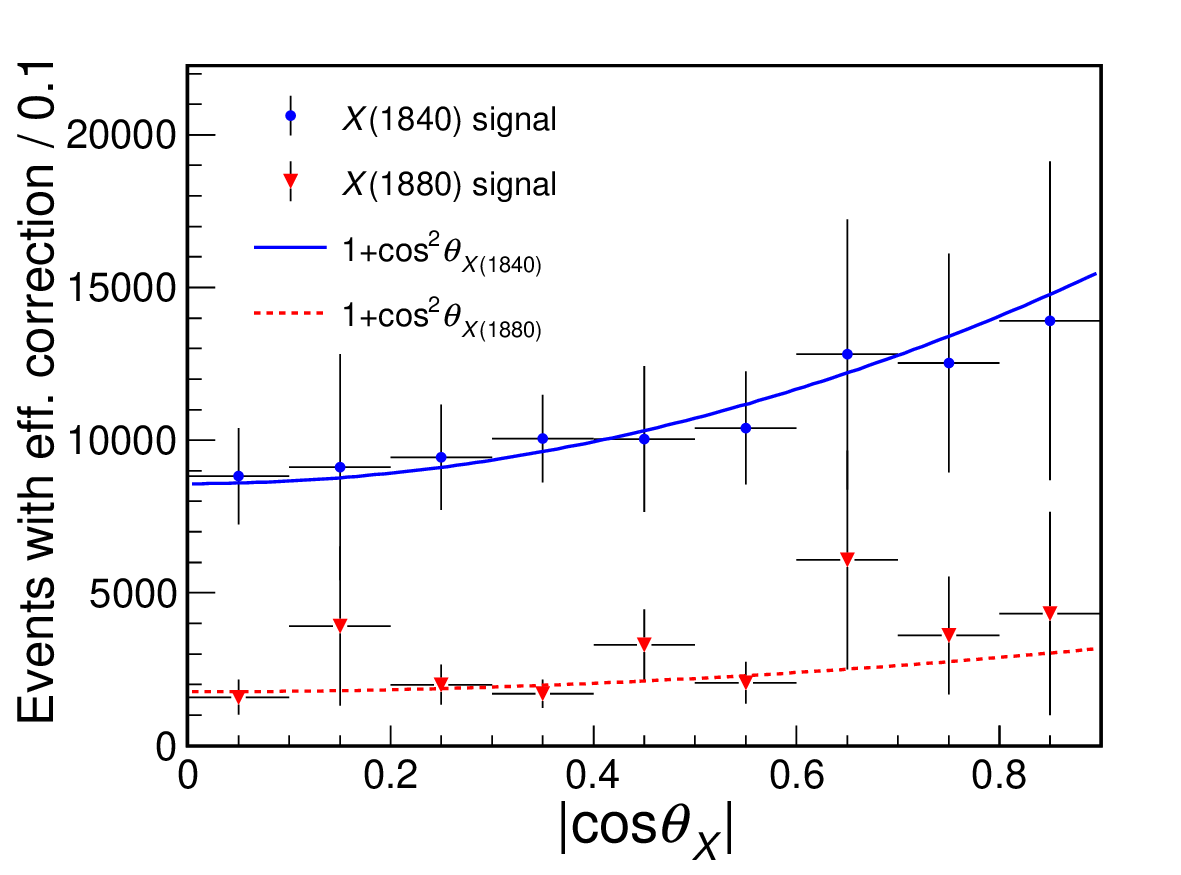}
    \caption{The acceptance-corrected angular distribution with respect to $\lvert \cos\theta \rvert$ for the $X(1840)$ and $X(1880)$, respectively. The curves represent the fit results with the function of $1+\cos^2\theta$ to the above two components: the solid curve in blue is for the $X(1840)$; the dashed one in red is for the $X(1880)$.\label{fig_angular}}
\end{figure}

Since the $X(1835)$ is known as a pseudoscalar meson, the $X(1840)$ is supposed to be a pseudoscalar particle as well considering the similar behaviours with those of the $X(1835)$. For the radiative $\jpsi$ decay to a pseudoscalar meson, the polar angle of the photon in the $\jpsi$ rest frame, denoted as $\theta$, is expected to follow a $1+\cos^2\theta$ distribution. 
The $\lvert \cos\theta \rvert$ is divided into nine bins in a region of $[0,0.9]$ to investigate the angular distribution.
The number of signal events corresponding to the constructive interference solution in each bin is obtained 
with the same fit procedure as mentioned above. The result is shown in Fig.~\ref{fig_angular}. As a result,
the angular distributions of $X(1840)$ and $X(1880)$ both agree with $1+\cos^2\theta$ and support the interpretation of the pseudoscalar mesons.
With the hypotheses of pseudoscalar mesons, the detection efficiencies for 
$J/\psi\rightarrow\gamma X(1840)$ and $J/\psi\rightarrow\gamma X(1880)$, 17.4\% and 18.4\%, are obtained from the MC simulation. The product branching fractions corresponding to that two solutions are summarized in Table ~\ref{table_finalresult}.

Sources of systematic uncertainties and their corresponding contributions to the measurement of the branching fractions are summarized in Table~\ref{table_err}. The uncertainties come from data-MC differences (tracking, photon detection,  4C kinematic fit etc.), total number of $\jpsi$ events, and background uncertainty from the change of fit range, MC model, MC statistics. For the MC model uncertainties due to the unknown spin-parity of the structures, we use the difference between phase space and a pseudoscalar meson hypothesis. In accordance with the previous publication~\cite{helix_corr}, we keep the efficiency with the track helix correction as the nominal value in this work, and take the difference between the efficiencies with and without correction as the systematic uncertainty from the 4C kinematic fit. The main contribution of systematic uncertainty comes from the uncertainty in the background estimation which is accessed by changing fit ranges. 
The uncertainty caused by the contribution above 2.07 GeV of the $M(6\pi)$ spectrum has a considerable effect on the  parameterization of the remaining background, which results in a large uncertainty of the branching fractions. Meanwhile, the impact of this uncertainty on the statistical significance of $X(1880)$ is considered, and the smallest statistical significance is chosen as the final result. The total systematic uncertainty is obtained by adding all of the mentioned ones in quadrature under the assumption that they are independent. The total systematic uncertainties on mass and width are estimated from the background uncertainty due to fit range and background description, and found to be $\pm\,2.5$ MeV/$c^2$ and $\pm\,7.7$ MeV for the $X(1840)$, $\pm\,0.7$ MeV/$c^2$ and $\pm\,2.4$ MeV for the $X(1880)$, respectively. Since the mass resolution of 4 MeV/$c^2$ is much smaller than the width of these structures, the uncertainty from the detector resolution is found to be negligible. 
The final results given in Table~\ref{table_finalresult} are obtained considering all the systematic uncertainties reported in Table~\ref{table_err}.

\begin{table}[htpb]
\begin{center}
\caption{The fitted parameters of the two coherent resonant structures and the corresponding product branching fractions. Solution I and II refer to the solutions characterizing model {\bf II} as discussed in the text.}
\label{table_finalresult}
\begin{ruledtabular}
\begin{tabular}{lcc}
	Parameters   &  Solution I & Solution II\\
	\hline
	$M_{X(1840)}$ (MeV/$c^2$) &\multicolumn{2}{c} {$1832.5\pm3.1\pm2.5$} \\
	$\Gamma_{X(1840)}$(MeV) & \multicolumn{2}{c} {$80.7\pm5.2\pm7.7$} \\	
	$\mathcal{B}_{X(1840)}(\times 10^{-5})$& $1.19\pm0.30\pm0.15$ &$2.07\pm0.50\pm0.36$  \\
	$M_{X(1880)}$ (MeV/$c^2$) & \multicolumn{2}{c} {$1882.1\pm1.7\pm0.7$} \\
	$\Gamma_{X(1880)}$(MeV) & \multicolumn{2}{c} {$30.7\pm5.5\pm2.4$}\\	
	$\mathcal{B}_{X(1880)}(\times 10^{-5})$&  $0.29\pm0.20\pm0.09$ &$1.19\pm0.31\pm0.18$ \\
\end{tabular}
\end{ruledtabular}
\end{center}
\end{table}

\begin{table}[htpb]
\begin{center}
\caption{Relative systematic uncertainties in the product branching fractions (in percent).}\label{table_err}
\begin{ruledtabular}
\begin{tabular}{ccc}
Sources       &     $X(1840)$ & $X(1880)$ \\
\hline
MDC tracking & 6.0 & 6.0 \\
Photon detection & 1.0 & 1.0 \\
Kinematic fit &  0.6 & 0.6\\
Detection efficiency &  3.0 & 3.1\\
Number of $\jpsi$ events & 0.4 & 0.4\\
\hline
Background uncertainty (Solution I) & 10.0 & 27.8 \\
Background uncertainty (Solution II) & 16.1 & 12.9 \\
\hline
Total (Solution I) & 12.1 & 28.6 \\
 Total (Solution II) & 17.5 & 14.6 \\
\end{tabular}
\end{ruledtabular}
\end{center}
\end{table}

In summary, a study of the radiative decay $\jpsi\ar\g3(\pip\pin)$ is performed with a sample of $(10087 \pm 44) \times 10^6$ $\jpsi$ events accumulated at the BESIII detector. A significant distortion of the $M(6\pi)$ distribution near the $p\bar{p}$ mass threshold is observed for the first time, which is analogous to the distortion observed in the $\pi^+\pi^-\eta^\prime$ invariant mass spectrum in $\jpsi\rightarrow\gamma\pip\pin\etap$~\cite{X1835_fit}.

To understand this anomalous line-shape, a few interpretations including a single structure described by a single BW or with threshold effect (model {\bf I}) and a coherent sum of two structures (model {\bf II}) are tested. We find that neither a simple BW nor a Flatt$\rm{\acute{e}}$ function could provide a reasonable description of data.
The scheme of a coherent sum of two structures gives a much better description on the anomalous line-shape in the $M(6\pi)$ spectrum. According to the fit results, the narrow structure, $X(1880)$, has a mass of $M=1882.1\pm1.7\pm0.7$ MeV/$c^2$ and a width of $\Gamma=30.7\pm5.5\pm2.4$ MeV. The significance of $X(1880)$ is larger than $10\sigma$ compared to the fit result with model {\bf I} considering any effect associated to the systematic uncertainties. The mass and width of $X(1840)$ are measured to be $M=1832.5 \pm 3.1 \pm2.5$ MeV/$c^2$ and $\Gamma=80.7\pm5.2 \pm7.7$ MeV, which are in agreement with the previous work~\cite{X1840_bes3_2013}.
Two solutions with the same fit quality and the identical resonant parameters but different branching fractions due to the constructive or destructive interference are summarized in Table~\ref{table_finalresult}.

Compared with the two structures observed in the $M(\pi^+\pi^-\eta^\prime)$ spectrum~\cite{X1835_fit}, the $X(1840)$ has a consistent mass with that $X(1835)$ but much narrower width.
The mass and width of the $X(1880)$ obtained in this work are in reasonable agreement with those reported in Ref.~\cite{X1835_fit}, which are $1870.2\pm2.2^{+2.3}_{-0.7}$ MeV/$c^2$ and $13.0\pm6.1^{+2.1}_{-3.8}$ MeV, respectively. This further supports the existence of a $p\bar{p}$ bound state just spanning the $p\bar{p}$ mass threshold.
At present, more sophisticated parameterizations such as a mixture of above two models cannot be ruled out. The observed anomalous line-shape in the $M(6\pi)$ spectrum in $\jpsi\ar\g3(\pip\pin)$ and the $\pi^+\pi^-\eta^\prime$ invariant mass spectrum in $\jpsi\rightarrow\gamma\pip\pin\etap$ reveal complex resonant structures near the $p\bar{p}$ mass threshold. To establish the relationship between different resonances in the mass region of $[1.8,1.9]$ GeV/$c^2$ and determine the nature of the underlying resonant structures, more data along with additional measurements including the determination of the spin-parity quantum numbers and the coupled channel amplitude analysis are highly desirable.

The BESIII Collaboration thanks the staff of BEPCII and the IHEP computing center for their strong support. This work is supported in part by National Key R\&D Program of China under Contracts Nos. 2020YFA0406300, 2020YFA0406400; National Natural Science Foundation of China (NSFC) under Contracts Nos. 11635010, 11735014, 11835012, 11905037, 11935015, 11935016, 11935018, 11961141012, 12022510, 12025502, 12035009, 12035013, 12061131003, 12192260, 12192261, 12192262, 12192263, 12192264, 12192265, 12221005, 12225509, 12235017; the Chinese Academy of Sciences (CAS) Large-Scale Scientific Facility Program; the CAS Center for Excellence in Particle Physics (CCEPP); Joint Large-Scale Scientific Facility Funds of the NSFC and CAS under Contract No. U1832207; CAS Key Research Program of Frontier Sciences under Contracts Nos. QYZDJ-SSW-SLH003, QYZDJ-SSW-SLH040; 100 Talents Program of CAS; The Institute of Nuclear and Particle Physics (INPAC) and Shanghai Key Laboratory for Particle Physics and Cosmology; ERC under Contract No. 758462; European Union's Horizon 2020 research and innovation programme under Marie Sklodowska-Curie grant agreement under Contract No. 894790; German Research Foundation DFG under Contracts Nos. 455635585, Collaborative Research Center CRC 1044, FOR5327, GRK 2149; Istituto Nazionale di Fisica Nucleare, Italy; Ministry of Development of Turkey under Contract No. DPT2006K-120470; National Research Foundation of Korea under Contract No. NRF-2022R1A2C1092335; National Science and Technology fund of Mongolia; National Science Research and Innovation Fund (NSRF) via the Program Management Unit for Human Resources \& Institutional Development, Research and Innovation of Thailand under Contract No. B16F640076; Polish National Science Centre under Contract No. 2019/35/O/ST2/02907; The Swedish Research Council; U. S. Department of Energy under Contract No. DE-FG02-05ER41374.

\bibliography{references}

\end{document}